\documentclass{iaus}
\usepackage{graphics}
\usepackage{natbib}
\newcommand{\arcm}{\hbox{$^\prime$}}
\newcommand{\arcs}{\arcm\hskip -0.1em\arcm}

\newcommand{\ie}{{\rm i.e.\ }}
\newcommand\aj{{AJ}}%
\newcommand\apj{{ApJ}}%
\newcommand\apjl{{ApJ}}%
\newcommand\apjs{{ApJS}}%
\newcommand\aap{{A\&A}}%
\newcommand\mnras{{MNRAS}}%
\newcommand\pasp{{PASP}}%
\newcommand\nat{{Nature}}%

  \checkfont{eurm10}
  \iffontfound
    \IfFileExists{upmath.sty}
      {\typeout{^^JFound AMS Euler Roman fonts on the system,
                   using the 'upmath' package.^^J}%
       \usepackage{upmath}}
      {\typeout{^^JFound AMS Euler Roman fonts on the system, but you
                   dont seem to have the}%
       \typeout{'upmath' package installed. iaus.cls can take advantage
                 of these fonts,^^Jif you use 'upmath' package.^^J}%
      }
  \else
  \fi

  \checkfont{msam10}
  \iffontfound
    \IfFileExists{amssymb.sty}
      {\typeout{^^JFound AMS Symbol fonts on the system, using the
                'amssymb' package.^^J}%
       \usepackage{amssymb}%

      }{}
  \fi

  \IfFileExists{amsbsy.sty}
    {\typeout{^^JFound the 'amsbsy' package on the system, using it.^^J}%
     \usepackage{amsbsy}}
    {}

\newsavebox{\astrutbox}
\sbox{\astrutbox}{\rule[-5pt]{0pt}{20pt}}

\newcommand\etal{\mbox{\textit{et al.}}}

\newcommand\eg{e.g.\ }

\title[Winds from Nuclear Starbursts]{Winds from Nuclear Starbursts: 
Old Truths and Recent Progress on Superwinds}

\author[David K.~Strickland]%
{David K.~Strickland}

\affiliation{
Dept.~of Physics \& Astronomy, Johns Hopkins University,
3400 N.~Charles St., Baltimore, MD 21218, USA.
email: dks@pha.jhu.edu}

\pubyear{2004}
\volume{222}
\pagerange{?--?}
\date{?? and in revised form ??}
\jname{The Interplay among Black Holes, Stars and ISM \\in Galactic Nuclei}
\editors{Th. Storchi Bergmann, L.C. Ho \& H.R. Schmitt, eds.}
\begin{document}

\maketitle

\begin{abstract}
I will discuss a few select aspects of the most common and
best understood galactic-scale outflow -- starburst-driven
superwinds, focusing on winds from nuclear starburst
galaxies. I will show that modern observations, in particular
in the soft and hard X-ray bands, complement
and reinforce the existing paradigm of superwinds as flows
collectively driven by multiple SNe. The properties of the diffuse
X-ray emission from dwarf starburst galaxies, 
$L_{\rm BOL} \sim L_{\star}$ starbursts in spiral galaxies,
and ULIRGS, are all consistent with superwind activity.
Where appropriate, I  contrast the physics of starburst-driven winds 
with poorly collimated winds from AGN, and
discuss what we know of the role of LLAGN and Seyfert nuclei in
starburst superwind galaxies.
\end{abstract}

\firstsection 
\section{Introduction}

It has long been appreciated that outflows from galaxies can have
a major effect on galaxy formation and evolution, influencing
such basic galactic properties such as the mean metal abundance
\citep{larson74,garnett02} or even the survival of low mass galaxies 
\citep{dekel86}. More
recent discoveries have reinvigorated interest in the role of galactic
outflows, e.g.: The presence of metals in the true
inter-galactic medium at low and
high redshifts \citep{songaila97,tripp00}; 
that a sizable fraction of all metals ever created now reside outside
galaxies \citep{pagel02}; 
and direct observational evidence for ubiquitous
outflows from, and 100 kpc-scale cavities
around, the Lyman break galaxies \citep{adelberger03}.

By the far the best-studied, best-understood, and arguably most
common form of galactic outflow capable of polluting the IGM are
superwinds  \citep{ham90,dahlem97}. 
These are loosely-collimated multi-phase outflows
from actively-star forming galaxies, \ie starburst galaxies.
Within the local universe starbursts account for 
$\sim 25$\% of all massive star formation 
(and hence metal production), and starburst activity becomes progressively
more important at higher redshifts \citep{heckman98}. 
As all local starburst galaxies appear to have 
superwinds \citep{lehnert96a}, it is clear that starburst-driven winds
are of major importance.

Over the last 5 years new observations of superwinds, in particular
satellite-based observations in the EUV and soft and hard X-ray bands,
have substantially added to our understanding of the physics behind
superwinds. I will discuss how these new observations
support the long-standing conceptual picture of how superwinds work,
specifically (a) how superwinds are driven by the collective
mechanical power of multiple supernovae (SNe) occurring within the disks
(in particular the nuclei) of starburst galaxies, and (b) X-ray
emission from superwinds. Where appropriate I
will contrast superwinds with AGN winds
(\ie loosely collimated outflows, and not large-scale jets).

\section{Distributed energy and mass injection in starburst nuclei}

The basic conceptual picture of how starburst-driven winds work
was established nearly two decades ago. In the 
\citet[henceforth CC]{chevclegg} model,
the very high rate per unit volume 
of core collapse SNe in starburst regions leads to young SNRs
colliding before they have time to lose energy radiatively. 
Shocks in these collisions thermalize the kinetic energy of the
ejecta, creating a very hot ($T\sim 10^{8}$ K),
high pressure ($P/k ~ \sim 10^{7}$ K cm$^{-3}$) and
low density gas that fills the volume of the starburst region.
In the analytical CC model the
multiple individual SN events are treated as a time-averaged,
spatially-uniform injection of mass and energy.
The merged SN-ejecta expands adiabatically, becoming supersonic 
at the edge of the starburst region and rapidly
reaching terminal velocity of $\surd(2\dot E/\dot M) \sim 3000$  km s$^{-1}$.
Numerical simulations
show that this wind expands preferentially along the minor axis of any
disk-like galaxy forming a bipolar flow, sweeping up 
cooler, denser, ambient disk or halo gas,
and stripping ambient gas from the walls of
the cavity \citep[\eg][]{ti88,suchkov94,ss2000}.
The ram pressure of the merged 
SN-ejecta accelerates this entrained gas to velocities of a few $\times100$ 
to 1000 km s$^{-1}$. The multiple cool, warm and hot gas phases observed in
superwinds \citep[\eg][]{dahlem97} are all entrained ambient gas.

Is this simple scenario realistic, given what we have learned
about starbursts in the last 20 years? Before the 
{\it Chandra} observations of \citet{griffiths2000} 
there had been no believable detection of the very hot gas
predicted by the CC model in any starburst 
region  (the very hot gas that in theory drives superwinds), so 
why was this model accepted
in the first place?
In fact, convincing observational evidence for the 
SN-driven nature of superwinds, and the quantitative
accuracy of the CC model, appeared soon after its publication. Modern
observations continue to validate its most basic concepts.

The classic starburst M82 is a particularly convenient laboratory
for investigating starbursts and superwinds, given that it is one
of the closest powerful starburst galaxies (D=3.6 Mpc), and
that all current SF occurs  within a 400 pc radius
of the nucleus. Its FIR luminosity, dominated by the
starburst activity, indicates a SN rate of $\sim 0.1$ per year.
In terms of SF rate per unit area M82 lies close to the 
upper limit found in starbursts at both low and high redshift
\citep{meurer97}.
Radio observations show  $\sim 40$ compact, but spatially resolved,
SNR remnants \citep{muxlow94}, that outline the 
starburst region as seen at other wavelengths. These are not the same SNe
that drive the wind, but are a subset of the SN population
that occur in dense gaseous environments,
although the total fraction of the starburst 
region filled with such dense gas must be $<0.1$ \citep{cf01}.
Optical, near and mid-IR observations show large numbers of
massive star clusters ($4 < \log M_{\rm cluster} < 6$) 
within the starburst region \citep{oconnell95,mccrady03,lipscy04}. Note 
that the fraction of massive stars (and hence eventual SNe)
within such massive clusters is only $\sim 20$\% of the total,
consistent with the value observed in many starburst galaxies
\citep{meurer95}. Thus, to first order, energy and mass injection
from SNe is distributed 
over the volume of the entire starburst region, in accord with the
CC model.

\citet{mccarthy87}, and later \citet{ham90} and \citet{lehnert96a}, 
used optical spectroscopy
to map the pressure in warm ionized gas in central regions of
M82 and other starburst galaxies with superwinds. Essentially this
uses the low filling factor H$\alpha$-emitting
 gas clouds as tracers of the much hotter
gas they are embedded in. At large radii the pressure drops 
rapidly with distance, as expected for a free radial wind.
Within a certain radius $r_{P}$ the pressure is approximately 
constant, indicating that energy and mass injection 
is distributed within this region. The radial shape 
and normalization of these pressure profiles are consistent with the CC model.
\citet{ham90} demonstrated for many galaxies with superwinds that
$r_{P}$ is the same as the size of starburst $r_{\star}$. 

Note that if  AGN drove these winds  $r_{P}$ would not equal $r_{\star}$, 
even in models where energy injection is distributed 
over large regions due to diverted or halted jets \citep{colbert97_thesis}.
A thermal AGN wind driven from the vicinity of the accretion disk
\citep[\eg][]{schiano85} would
further differ from a starburst-driven wind, even if having initially
similar total mass and energy injection rates. 
The CC model assumes radiative losses from the thermalized 
gas are negligible, which is true if energy and mass
injection occurs over a volume similar to the size of an entire starburst 
region. \citet{silich03} have pointed out that 
if the injection region is much smaller than this, 
the increased density of the thermalized gas leads to significant
radiative losses. Their model is essentially of flows from
very massive ($\log M > 6$) super star clusters, but I wish to
point out that it would also apply to centrally-driven AGN winds. 
Strong radiative losses in AGN winds would thus alter the
physical properties observed on the larger galactic scales from those
in starburst-driven superwinds.

\begin{figure}
\centerline{
\scalebox{0.7}{%
\includegraphics{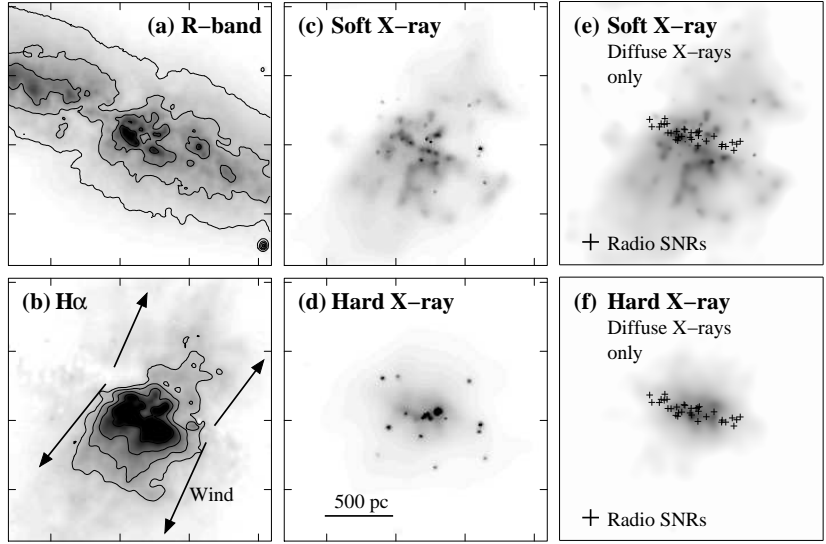}%
}
}
\caption{The center of M82 in (a) the optical R-band,
(b) H$\alpha$ emission from $\log T \sim 4$ gas, (c) Soft X-ray emission
	(E=0.3-2.0 keV), (d) Hard X-ray emission (E=2.0-8.0 keV),
	(e) Diffuse soft X-ray emission from $\log T \sim 6.5$ gas 
	only, X-ray point sources have
	been removed, (f) Diffuse hard X-ray emission, including some
	contribution from $\log T \sim 7.5$ -- 8 gas.  
	Each image is 2 kpc on a side, centered
	on the dynamical center of the galaxy. X-ray images are
	from {\it Chandra} ACIS observations. The location of the $\sim40$
	compact radio SNRs that mark the starburst region
	are plotted on panels e and f. The base of the 
	superwind matches the size of the starburst region. Adapted
	from \citet{strickland04a}.}
\label{fig:dks_fig1}
\end{figure}

Recent X-ray observations of superwinds have validated the use of warm
gas as a tracer of the hot merged SN-ejecta, 
as well as detecting the very hot
plasma predicted by the CC model. It is only
with the {\it Chandra} X-ray Observatory's arcsecond spatial 
resolution that emission from the multiple X-ray binaries 
formed in the starburst can be cleanly  separated from
the diffuse X-ray emission associated with the superwind 
(see Fig.~\ref{fig:dks_fig1}).
This is especially true in the hard X-ray band ($E=2$ to 8 keV)
where \emph{faint} diffuse bremsstrahlung 
and line emission from $T\sim10^{8}$ K gas would be expected
to be seen. \citet{griffiths2000} were the first to robustly
detect diffuse hard X-ray emission in a starburst region, once
again in the archetype M82.
The spatial extent of this emission in the plane of the galaxy
exactly matches that of the $40\arcs$-diameter 
starburst region, and matches the
base of the superwind as seen at other wavelengths.
More recent, higher spectral-resolution, {\it Chandra} 
observations (Strickland
et.~al, in preparation) show that the diffuse hard X-ray emission 
not purely thermal, but do confirm the presence of a $E\approx6.7$ 
keV Fe K-shell emission that must come from a
$\log T > 7.5$ K plasma. In general such diffuse hard 
emission has not been detected in other nearby starbursts, even
with {\it Chandra}, as other starburst nuclei are
more distant, and have smaller angular scales, than 
that in M82. \citet{weaver02} detected diffuse hard X-ray emission
within the central $\sim10\arcs$ of the nuclear starburst NGC 253,
but argue that the properties of this emission are more consistent with
gas photoionized by that galaxy's LLAGN.

\section{Soft X-ray emission from superwinds: theory vs observation}

Diffuse soft X-ray emission in superwinds is considerably easier to
observe than diffuse hard X-ray emission, 
by virtue of being orders of magnitude more
luminous and spatially extended. Nevertheless {\it Chandra} has also
revolutionized our understanding of the origin and
properties of soft diffuse X-ray emission in starbursts.
The majority of starburst activity in the local universe occurs
in galaxies with $\log L_{\rm BOL} (L_{\odot}) \sim 10.5$, \eg classic
starbursts M82, NGC 253, or NGC 3628. Thus edge-on starburst
galaxies such as these objects, at distances of $\lesssim 20$ Mpc,
provide the best observational data on superwinds.

Models such as Starburst99 \citep{leitherer99} predict that
the ratio of mechanical energy released by SNe and stellar winds
to the bolometric radiated power from a starburst is 
$L_{\rm W}/L_{\rm BOL} \approx 0.01$ to 0.015, where
$L_{\rm BOL}$ is proportional to the star formation rate.
The fraction of the superwind mechanical power expected to emerge as soft 
X-ray emission is rather more uncertain. There are a variety of
different theoretical models for the origin of the soft X-ray emitting
gas.
The merged SN ejecta driving the wind
is too tenuous to provide significant X-ray emission, unless its
density has been significantly increased by mass-loading 
\citep[\eg][]{suchkov96}.
In our multi-dimensional
hydrodynamical simulations of superwinds \citep{ss2000}, where 
the emission is primarily due to shock heating of ambient 
disk and halo gas,
we found $L_{\rm X}/L_{\rm W} \sim 0.03$, 
(with higher values associated with mass-loaded models).
Similar values can be found if the \citet{weaver77} wind-blown bubble
model is applied to superwinds \citep[\eg][]{strickland04b}, 
where the soft X-ray emission is primarily from
conductively-heated ambient gas at the wall of the bubble. 
Combining these models, and
including the scatter, theoretical models of
starburst-driven winds predict
$\log L_{\rm X}/L_{\rm BOL} \approx -3.5\pm{0.6}$.

{\it Chandra} observations of many classic starbursts
have now demonstrated that the soft thermal
X-ray emission from the superwinds is highly-structured and
spatially correlated with 
H$\alpha$ emission, as expected if the X-ray emission is
generated by the interaction of the superwind with cool ambient
gas. In many cases both the X-ray and H$\alpha$ emission is strongly
limb-brightened, indicating that the emission comes predominantly
from the walls
of the cavity carved by the superwind into the disk and halo medium
of the host galaxy \citep{strickland00,strickland02,strickland04a}. 
The fraction of the wind volume occupied by
soft X-ray emitting gas is relatively low, with upper limits of
$< 20$\% of the soft X-ray emission coming from any volume-filling
mass-loaded wind \citep{strickland00,strickland03}. Both the total
diffuse soft X-ray luminosity $L_{\rm X,TOT}$ 
and the superwind emission at heights $z> 2$ kpc
from the disk $L_{\rm X,HALO}$, scale in direct proportion to
the the host galaxies bolometric luminosity, 
$\log L_{\rm X,TOT}/L_{\rm BOL} \approx -3.6\pm{0.2}$ 
and $\log L_{\rm X,HALO}/L_{\rm BOL} \approx -4.4\pm{0.2}$  
\citep{strickland04b}, remarkably tight correlations.



Chandra and XMM-Newton observations of many dwarf 
starbursts ($L_{\rm BOL} \lesssim 9$)
and ULIRGs ($\log L_{\rm BOL} \gtrsim 12$) are 
now available \citep{martin02,ott03,ptak03,summers03,summers04,hartwell04}. 
The characteristics of the diffuse X-ray emission
in these objects are essentially the same as those of the
classic starbursts, with similar spectral properties (\eg relative
$\alpha$-to-Fe abundances) and $L_{\rm X}/L_{\rm BOL}$ ratios
(Grimes \etal, in preparation). 

\section{Outflows from galaxies with both starburst and AGN activity}

The soft X-ray properties of many actively
star-forming galaxies with
spatially-extended soft X-ray emission are in accord with models
of supernova-driven winds. But what of starburst galaxies that
also host AGN? In general the outflows from classic 
starburst galaxies that also host LLAGN (\eg NGC 253, NGC 3079) 
do not differ from 
winds from pure starbursts (\eg M82, NGC 3628, see \citealt{strickland04a}). 
The one ``classic'' starburst that does appear unusual
is NGC 4945, which is also the host of a very peculiar X-ray-luminous
AGN that is probably heavily
obscured along all lines of sight \citep{marconi00,levenson02}. 
Although it has a X-ray and H$\alpha$
nuclear outflow cone similar to NGC 253, it is lacking in
diffuse X-ray or H$\alpha$ emission when compared to a starburst
of the same \emph{total} galactic bolometric luminosity,
$\log L_{\rm X,TOT}/L_{\rm BOL} = -4.25$ \citep{strickland04a}. 
However, the AGN may dominate
$L_{\rm BOL}$. If we assume that the diffuse X-ray emission is due to
a starburst-driven wind alone, then the starburst must only account
for $\lesssim 20$\% of $L_{\rm BOL}$. In general, the large-scale soft
X-ray emission in Seyfert/Starburst composite galaxies is consistent
with a purely-starburst origin \citep{levenson01a,levenson01b,levenson04}. 
Thus, for a given total bolometric luminosity it appears that SNe are more
effective at driving galactic-scale winds than AGN..
This does not imply that AGN-driven galactic winds do not exist. 
There clearly are galaxies with AGN but lacking starbursts 
that have galactic-scale (\ie $\sim 10$ kpc) outflows 
\citep{colbert96b,colbert98}, but
their local space density is lower than 
typical starburst superwind galaxies.

%
%

\section{Final remarks}

Many questions remain to be solved regarding
the physics within superwinds, and their larger
cosmological impact on galaxies and the IGM.
Nevertheless, we have a good physical understanding of how SNe within
starbursts drive galactic-scale superwinds. 
The standard superwind conceptual model is almost two decades old, 
but modern observation validate it far more often than not.
Future research in this area, in particular the transition between 
starburst activity (with superwinds)  and ``normal'' 
galaxies (with or without galactic fountains), will strengthen
our understanding of the interplay between SF, stars, and the ISM
in regulating galaxy assembly, structure and evolution.

I would like to thank my principal collaborators on
the work mentioned in this contribution:
E.~Colbert, M.~Dahlem,
T.~Heckman, C.~Hoopes, I.~Stevens and K.~Weaver.
This work has been supported by grants from the {\it Chandra}
Fellowship and GO programs.


\end{document}